\documentclass[11pt]{article}
\usepackage{moriond,epsfig}

\def\Journal#1#2#3#4{{#1} {\bf #2}, #3 (#4)}

\def\PRL{\em Phys. Rev. Lett.}
\def\PRD{{\em Phys. Rev.} {\bf D}$\!$}

\newcommand{\Eq}[1]{Eq.~(\ref{eq#1})}
\def\st{\scriptstyle}

\def\dm#1{\Delta m^2_{\mathrm{\st #1}}}
\def\lsim{\;\raisebox{-.6ex}{$\stackrel{<}{\sim}$}\;}

\newcommand{\optbar}[1]{\shortstack{{\tiny (\rule[.4ex]{1em}{.1mm})} 
  \\ [-.7ex] $#1$}}
\def\qsin2b#1{\left. ``\sin 2\beta \mbox{\,\textquotedblright} \right| _{\mathrm{From}\:#1 K_S} }
\def\tchi{\tilde{\chi}_1^0}
\def\be{\begin{equation}}
\def\ee{\end{equation}}
\def\bea{\begin{eqnarray}}
\def\eea{\end{eqnarray}}

\def\ra{\rightarrow}
\def\epp{\epsilon^{\prime}}
\def\etal{{\em et al.}}
\begin{document}
\vspace*{4cm}
\title{SUMMARY OF THE 2003 MORIOND WORKSHOP ON ELECTROWEAK INTERACTIONS AND UNIFIED THEORIES\footnote{To appear in the Proceedings of the XXXVIIIth Rencontres de Moriond on Electroweak Interactions and Unified Theories (held March, 2003), edited by J. Tr\^an Thanh V\^an.}}

\author{ BORIS KAYSER}

\address{Fermi National Accelerator Laboratory \\
P.O. Box 500, Batavia IL 60510}

\maketitle
\abstracts{
We recount some of the highlights of the 2003 Moriond Workshop on Electroweak Interactions and Unified Theories.}

\section{Introduction}

The 2003 Moriond Workshop on Electroweak Interactions and Unified
Theories covered a very rich, diverse array of recent results
concerning neutrinos, astrophysics and cosmology, searches for new
particles, Higgs physics, precision low-energy measurements, quark
flavor physics, CP violation, and electroweak interactions. In this
summary, we recapitulate some of the highlights.

We update many of the results reported at the Workshop to include
newer findings reported during Summer, 2003.

\section{Neutrinos}

Neutrinos are among the most abundant particles in the universe. Thus,
to understand the universe, we must understand the neutrinos. Recently,
our comprehension of their nature has increased dramatically. We have
obtained evidence, whose great strength is summarized in Table \ref{t1}, that
neutrinos can change from one flavor to another. Barring exotic
possibilities, neutrino flavor change implies neutrino mass and
mixing. Thus, we have learned that neutrinos almost certainly have
nonzero masses and mix.
\begin{table}[ht]
\caption{The strength of the evidence for neutrino flavor change. The symbol $L$ denotes the distance travelled by the neutrinos.
\label{t1}} \vspace{0.1in}
\begin{center}
\begin{tabular}{@{}ll}
 \hline
Neutrinos & Evidence for Flavor Change\\
\hline
Atmospheric & Compelling\\
Accelerator ($L=250\,$km) & Interesting\\
Solar & Compelling\\
Reactor ($L\sim 180\,$km) & Very Strong\\
From Stopped $\mu^+$ Decay (LSND) & Unconfirmed\\
\hline
\end{tabular}
\end{center}
\end{table}

Particularly compelling evidence that {\em solar} neutrinos change
flavor has been reported by the Sudbury Neutrino Observatory
(SNO).~\cite{r1} SNO measures the high-energy part of the solar
neutrino flux arriving at earth using three different detection
reactions. As summarized in Table \ref{t2},
\begin{table}[ht]
\caption{The detection reactions employed by SNO, and the fluxes they measure.
\label{t2}} \vspace{0.1in}
\begin{center}
\begin{tabular}{cc}
\hline \vspace{0.1cm}
\underline{Detection Reaction}    & \underline{Flux Measured} \\
$\nu d \to \nu np$  & $\phi_e + \phantom{0.15\,}\phi_{\mu\tau}$ \\
$\nu e \to \nu e$    & $\phi_e + {0.15\,}\phi_{\mu\tau}$ \\
$\nu d \to epp$        & $\phi_e   \phantom{+ 0.15\,\phi_{\mu\tau}}$ \\
\hline
\end{tabular}
\end{center}
\end{table}
the observed rates of these reactions determine three different linear
combinations of the arriving solar $\nu_e$ flux, $\phi_e$, and the
$\nu_\mu + \nu_\tau$ flux, $\phi_{\mu\tau}$. From the observed rates
for the two deuteron breakup reactions, SNO finds that~\cite{r2}
\be
\frac{\phi_e}{\phi_e + \phi_{\mu\tau}} = 0.306 \pm 0.026 \,\mathrm{ (stat)} 
\pm 0.024 \,\mathrm{ (syst)}~~.
\label{eq1}
\ee
Clearly, the flux $\phi_{\mu\tau}$ of muon and/or tau neutrinos
ariving from the sun is nonzero. But all the solar neutrinos are born
in nuclear reactions that produce only electron neutrinos. Hence,
neutrinos obviously do change flavor.

Corroborating information comes from the detection reaction $\nu e \ra
\nu e$, studied by both SNO and Super-Kamiokande (SK).~\cite{r3}

The strongly favored explanation of solar neutrino flavor change is
the Large Mixing Angle version of the Mikheyev Smirnov Wolfenstein
Effect (LMA-MSW). This interpretation implies that by the time reactor
$\overline{\nu_e}$ have traveled $\sim$ 200 km, a substantial fraction of
them should have disappeared into antineutrinos of other flavors. Very
interestingly, the KamLAND experiment confirms that reactor $\overline{\nu_e}$ do indeed disappear. The $\overline{\nu_e}$ studied by KamLAND have typically
travelled $\sim$180 km, and KamLAND finds that the $\overline{\nu_e}$ flux
is only $0.611 \pm 0.085 \pm 0.041$ of what it would be if none of it
were disappearing.~\cite{r4} Both this reactor $\overline{\nu_e}$
disappearance and the solar neutrino results can be described by the
same neutrino mass and mixing parameters,~\cite{r5,r2} bolstering one's
confidence that the physics of both phenomena has been correctly
identified. The successful parameters include a neutrino (mass)$^2$
splitting $\dm{sol} \sim 7 \times 10^{-5} \,\mathrm{eV}^2$, and a mixing angle $\theta_{\mathrm{\st sol}} \sim 32^\circ$.~\cite{r2}\

Compelling evidence that {\em atmospheric} neutrinos change flavor has
come from voluminous data on the interactions of these neutrinos in
and beneath underground detectors. These data are beautifully
described, in detail,~\cite{r6} by the neutrino flavor oscillation
$\nu_\mu \ra \nu_\tau$ with a neutrino (mass)$^2$ splitting $\dm{atm}
\simeq 2.0 \times 10^{-3}\,\mathrm{eV}^2$,~\cite{r7} and large (possibly
maximal) mixing.

The K2K experiment \cite{r8} seeks to confirm the atmospheric neutrino
flavor change by showing that {\em accelerator} neutrinos undergo this
change too. In K2K, an accelerator $\nu_\mu$ beam passes through a
near-detector complex that determines the initial $\nu_\mu$ flux. The
beam then travels 250 km, and passes through a far detector that
determines what fraction of the initial $\nu_\mu$ flux is still
present. From the near-detector measurements, 80 $\nu_\mu$ events were
expected in the far detector, assuming no disappearance of $\nu_\mu$
into other flavors, but only 56 events were seen. The K2K data are
well described by the same oscillation hypothesis that describes the
atmospheric neutrino data, with the same parameters.~\cite{r8,r9}

Subsequent to the Moriond Workshop, K2K reported that in a new data
sample, 26 $\nu_\mu$ events were expected in the absence of
disappearance, but only 16 events were seen.~\cite{r10} This degree of
disappearance is consistent with that seen earlier.

So-far unconfirmed evidence for short-wavelength $\overline{\nu_\mu} \ra
\overline{\nu_e}$ oscillation with a large (mass)$^2$ splitting $\dm{LSND}
\sim 1 \,\mathrm{eV}^2$ has come from the LSND experiment.~\cite{r11} If
LSND is confirmed, we will have three quite different (mass)$^2$
splittings: $\dm{sol} \sim 7 \times 10^{-5} \,\mathrm{ eV}^2$, $\dm{atm}
\sim 2 \times 10^{-3} \,\mathrm{ eV}^2$, and $\dm{LSND} \sim 1 \,\mathrm{ eV}^2$.
Assuming CPT invariance and taking other data into account, this set
of (mass)$^2$ splittings will imply that nature contains at least one
``sterile'' (non-weakly-interacting) neutrino. Thus, it is of great
interest to confirm or refute LSND, and the currently running MiniBooNE experiment will do that.~\cite{r12}

\begin{figure}[htb]
\begin{center}
\psfig{figure=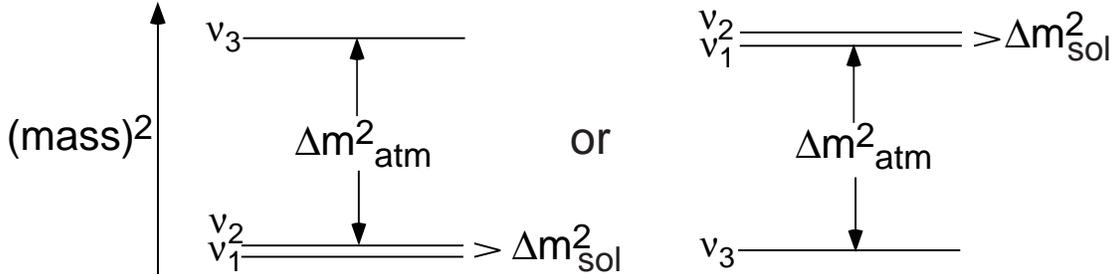,height=1.5in}
\caption{The neutrino (mass)$^2$ spectrum assuming only three neutrinos.}
\label{f0}
\end{center}
\end{figure}

If MiniBooNE does not confirm LSND, then nature may contain just three neutrino mass eigenstates, $\nu_1,\: \nu_2$, and $\nu_3$. The (mass)$^2$ spectrum is then as depicted in Fig.~\ref{f0}. 
Surprisingly, each mass eigenstate couples substantially to more than one charged lepton flavor. This fact is reflected in a leptonic mixing matrix $U$ that contains two {\em large} mixing angles. This matrix has the character
\begin{equation}  
\begin{array}{c} \phantom{x} \\ U \sim \end{array} \begin{array}{cc}
	\hspace{-0.25in}\nu_1  \hbox{\hskip0.3cm} \nu_2  \hbox{\hskip0.3cm} 
	\nu_3 \\
\left[  \begin{array}{ccc}
	B & B & s \\
	B & B & B \\
	B & B & B 
\end{array}  \right]
\begin{array}{c}
 	e  \\  \mu \\ \tau 
\end{array}  \end{array} ~~.
\label{eq2} 
\end{equation}
Here, each column corresponds to the neutrino mass eigenstate listed above it, and each row to the charged lepton listed to its right. The symbol $B$ denotes a real matrix element that is {\Large big} (i.e., not negligible compared to unity), while $s$ is the sole {\footnotesize small} element and is given by 
\be
s = e^{-i\delta} \sin \theta_{13}~~.
\label{eq3}
\ee
Here, $\delta$ is a CP-violating phase and $\theta_{13}$ is the only mixing angle in $U$ that is relatively small. At present, we know only that, at $3 \sigma, \:\theta_{13} \lsim 15^\circ$. \cite{r13}

The phase $\delta$, if present, will lead to the CP-violating inequality
\be
P(\nu_\alpha \ra \nu_\beta) \neq P(\overline{\nu_\alpha} \ra \overline{\nu_\beta})
\label{eq4}
\ee
between the probability $P(\nu_\alpha \ra \nu_\beta)$ for a neutrino of flavor $\alpha\;(=e,\,\mu$, or $\tau)$ to oscillate into one of flavor $\beta$, and the probability for the corresponding antineutrino oscillation. Observing such an inequality would establish that CP violation is not a peculiarity of quarks, but also occurs among leptons.

As we see from \Eq{3}, the phase factor $e^{-i\delta}$ enters the matrix $U$ multiplied by $\sin \theta_{13}$. Thus, the size of the CP-violating asymmetries produced by $\delta$ depends on the value of $\theta_{13}$. Consequently, what experimental facilities we will need to see and study these asymmetries will depend on whether this value is only somewhat smaller, or much smaller, than the current bound. At this workshop, M. Mezzetto discussed the range of $\theta_{13}$ for which CP violation would be visible in a superbeam (a conventionally generated, but intense, $\optbar{\nu_\mu}$ beam), a beta beam (a $\optbar{\nu_e}$ beam generated by the decay of radioactive nuclei), a neutrino factory beam (produced by the decay of stored muons), or by using combinations of these facilities. \cite{r14} In addition, in parallel with this workshop, there was another focussed on the uses of beams of radioactive nuclei. \cite{r15}

Our ultimate goal in the neutrino realm is to understand what physics is responsible for neutrino masses and mixing. One challenge is to understand why two of the leptonic mixing angles are so large compared to their quark counterparts. \cite{r16} Another is to understand why the neutrino masses are so small compared to their quark counterparts. Perhaps this smallness results from the fact that non-interacting neutrino fields can escape into extra spatial dimensions. The upper bound on the size of such extra dimensions coming from limits on supernova neutrinos escaping into them is now $\sim 10^{-(5-6)}$ cm, not as tight as had been previously thought. \cite{r17}

To uncover the physics behind neutrino mass and mixing, we will first have to gain a much better knowledge of the neutrino properties and parameters. Present experimental results allow a wide range of possilbe values for quantities to be determined in the future. \cite{r18,r19} In general, future measurements will not determine these quantities one at a time, but will yield combinations of them. A disentanglement of parameters will then be necessary. Promising strategies for determining individual neutrino parameters through complementary measurements at different experimental facilities are being developed. \cite{r20}

Some of the important open questions about neutrinos will be addressed by experiments using accelerator neutrino beams, while others will require non-accelerator experiments. Among the latter are the question of whether neutrinos are their own antiparticles, which can be addressed by searching for neutrinoless nuclear double beta decay, and the question of the absolute scale of neutrino mass, which can be attacked via experiments on single and double nuclear beta decay, and perhaps through cosmological observations. A major new tritium beta decay experiment is being mounted, and numerous new, more sensitive searches for double beta decay are being considered, \cite{r21} or are already running. \cite{r22}

\section{Leptogenesis}

As Sakharov pointed out long ago, the present baryon asymmetry of the universe could not have developed without a suitable violation of CP at some point in the universe's evolution. Interestingly, it has been found that CP violation from the one established source --- the CP-violating phase in the quark mixing matrix --- would not have been sufficient. As a result, there is great interest in the possibility that the baryon asymmetry arose through leptogenesis.~ \cite{r23}$^-$\cite{r27} The most popular scenario for leptogenesis is a natural outgrouth of the appealing ``see-saw'' explanation of why neutrinos are so light. In the see-saw picture, the light neutrinos have very heavy Majorana neutral partners, $N$. These particles $N$ would have been produced in the Big Bang. If their decays into charged leptons $\ell$ violate CP, then we can have the decay rate inequality
\be
\Gamma (N \ra \ell^+ + \dots) \neq \Gamma (N \ra \ell^- + \dots)~~,
\label{eq5}
\ee
which would have led to unequal numbers of $\ell^+$ and $\ell^-$ in the universe. This leptogenesis would have been followed by nonperturbative Standard Model processes that would have converted the lepton asymmetry, in part, into our baryon asymmetry.

An obvious question is how the CP violation required for leptogenesis via the unequal heavy rates of \Eq{5} is related to the CP violation that we hope to observe in neutrino oscillation today. The answer is that, while the relation is model-dependent, it is not likely that we have one of these violations of CP without the other.~\cite{r23}$^-$\cite{r31} This certainly heightens the interest in observing CP violation in neutrino oscillation.

For the leptogenesis scenario we have sketched to lead to a baryon asymmetry of the observed magnitude, the mass $M_1$ of the lightest $N$ must exceed $3\times 10^8$ GeV. On the other hand, in supersymmetry, we must have $M_1 < 10^{(9-12)}$ GeV. \cite{r24} Thus, if supersymmetry is found, the range of possible $M_1$ values for which leptogenesis can work will have been narrowed.

\section{Dark Matter}

The quarks, charged leptons, and neutrinos provide but a small fraction of the mass density of the universe. Some other form of matter --- Dark Matter that does not shine --- provides a much larger fraction. What is this Dark Matter? Perhaps it consists of Weakly Interacting Massive Particles (WIMPS), such as the lightest neutralino, $\tchi$, in popular supersymmetric theories. \cite{r32} Bounds from terrestrial WIMP searches appear to exclude the DAMA signal reported earlier, but future searches will be sensitive to a much larger region of model parameter space. \cite{r33,r34}

In supergravity-inspired supersymmetric models, if the Lightest Supersymmetric Particle (LSP) is the neutralino $\tchi$, and it is stable, then its naively predicted relic abundance in the universe is much too high, and it provides a Dark Matter density much greater than what is observed.~\cite{r46} However, annihilation of $\tchi$ against slightly heavier supersymmetric particles could lead to a lower relic $\tchi$ abundance compatible with observation.~\cite{r46}

\section{Observational Cosmology}

Extremely impressive information on the universe is coming from new observations. At this workshop, we heard about new results on the cosmic microwave background fluctuation power spectrum. \cite{r35,r36} Analysis \cite{r37} of these and other cosmological data yields for the total energy density in the universe, relative to the critical density for closure, a value $\Omega_{\mathrm{tot}} = 1.02 \pm 0.02$, for the fraction of this which is matter a value $\Omega_M = 0.27 \pm 0.04$, and for the fraction which is Dark Energy a value $\Omega_\Lambda = 0.73 \pm 0.04$. \cite{r37,r35}

\section{Searches for SUSY and Other New Particles}

The existence of Dark Matter of unknown composition is one more reason to look for new particles, such as those predicted by supersymmetry, at colliders. So far, the searches, a number of which were reported at the workshop, \cite{r38}$^-$\cite{r44} have been mostly negative. For example, LEP searches have excluded, at 95\% C.L., a stable $\tilde{t}$ quark with mass less than 95 GeV. \cite{r39} However, some experiments have yielded tantalizing results. \cite{r44} At HERA, the H1 experiment sees an anomalous number of events with an isolated electron or muon and missing transverse momentum, $p_{T,\:\mathrm{miss}}$. For $p_{T,\:\mathrm{miss}} > 25$ GeV, 10 events are seen while only $2.9 \pm 0.5$ would have been expected from the Standard Model. For $p_{T,\:\mathrm{miss}} > 40$ GeV, 6 events are seen while only $1.1 \pm 0.2$ would have been expected. There is no excess of events with an isolated $e$ or $\mu$ in the ZEUS experiment, but ZEUS sees an anomalous number of events with an isolated $\tau$. For $p_{T,\:\mathrm{miss}} >25$ GeV, 2 events are seen where $0.12 \pm 0.02$ would have been expected from the SM. For $p_{T,\:\mathrm{miss}} > 40$ GeV, 1 is seen where $0.06 \pm 0.01$ would have been expected. This is certainly an interesting situation. (The numbers reported at Moriond were reported again at Lepton Photon 2003, five months later. \cite{r45})

\section{Studying the Standard-Model Higgs}

If a likely candidate for the Standard-Model (SM) Higgs boson $H$ is discovered, we will want to confirm that it is indeed the Higgs particle. We may do that by verifying, among other things, that it has a signature property of $H$ --- a coupling to any fermion that is proportional to the fermion's mass. This property implies that, in particular, the coupling of $H$ to the top quark is large.

The cross sections for $p\bar{p} \ra t\bar{t}H$ at the Tevatron and for $pp \ra t\bar{t}H$ at the LHC have been calculated at ${\cal O}(\alpha_S^3)$. \cite{r47} The $t\bar{t}H$ final state appears to be a very interesting one, both as a Higgs boson discovery mode and as a way to confirm that a candidate Higgs does have the signature Higgs couplings.

\section{New Physics Through Precision}

\begin{figure}[htb]
\begin{center}
\psfig{figure=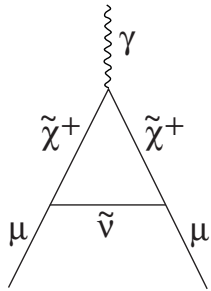,height=1.5in}
\caption{A non-Standard-Model contribution, coming from supersymmetry, to the gyromagnetic ratio of the muon. The particle $\tilde{\chi}^+$ is a chargino, and $\tilde{\nu}$ a sneutrino. 
\label{f1}}
\end{center}
\end{figure}

Very precise measurements are being made of $a_\mu \equiv (g_\mu - 2) / 2$, the anomalous part of the gyromagnetic ratio, $g_\mu$, of the muon. If the measured value of $a_\mu$ should disagree definitively with the SM prediction, we will
have evidence for New Physics. An example of the non-SM New Physics that could contribute to $a_\mu$ is shown in Fig.~\ref{f1}.

\begin{figure}[hbt]
\begin{center}
\psfig{figure=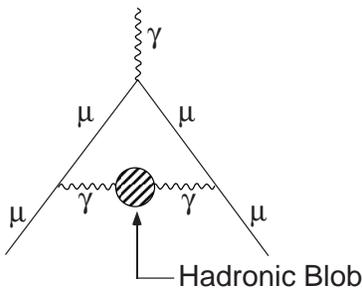,height=1.5in}
\caption{The hadronic contribution to the vacuum polarization. 
\label{f2}}
\end{center}
\end{figure}

The major uncertainty in the SM prediction for $a_\mu$ is the hadronic contribution to the vacuum polarization, depicted in Fig.~\ref{f2}. 
\cite{r48,r49} Within this process, Amp $[\gamma \ra$ Hadronic Blob $\ra \gamma] \sim$ Amp$[\gamma \ra$ Hadrons$] \,\times$ Amp[Hadrons $\ra \gamma] \sim |$Amp$[\gamma \ra$ Hadrons]$|^2$. The latter can be determined by measuring $\sigma [e^+ e^- \ra \gamma \ra$ Hadrons], and factoring out the known coupling of an $e^+e^-$ pair to a photon. \cite{r48,r49}

\begin{figure}[htb]
\begin{center}
\psfig{figure=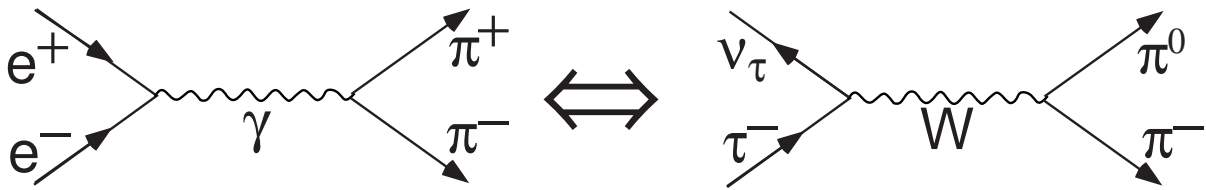,width=16cm}
\caption{Two processes related by CVC.
\label{f3}}
\end{center}
\end{figure}

An alternate way to determine the contribution of the Hadronic Blob, at least of its important $\pi^+\pi^-$ component, is to note that, by CVC, , the amplitude for $e^+ e^- \ra \gamma \ra \pi^+\pi^-$ can be inferred from that for $\tau^- \ra \nu_\tau + \pi^0\pi^-$ (cf. Fig.~\ref{f3}). \cite{r48,r49} Thus, one may use $\tau$ decay data in place of data on $e^+ e^- \ra$ Hadrons.

The experimental value of $a_\mu$ is \cite{r50}
\be
a_\mu^{\exp} = (11\,659\,203 \pm 8) \times 10^{-10}~~.
\label{eq6}
\ee
Midway through the Moriond workshop, it was reported \cite{r49,r48} that if one uses the $e^+ e^-$ data to determine the hadronic vacuum polarization contribution, then the consequent SM prediction for $a_\mu,\; a_\mu^{SM}(e^+ e^-)$, disagrees with the measured value, $a_\mu^{\exp}$, by
\be
a_\mu^{\exp} - a_\mu^{SM}(e^+ e^-) = (35.6 \pm 11.7) \times 10^{-10}~~,
\label{eq7}
\ee
a $3\sigma$ discrepancy. On the other hand, if one uses the $\tau$ decay data, then the consequent SM prediction for $a_\mu,\; a_\mu^{SM}(\tau)$, differs from $a_\mu^{\exp}$ by \cite{r49,r48}
\be
a_\mu^{\exp} - a_\mu^{SM}(\tau) = (10.4 \pm 10.7) \times 10^{-10}~~,
\label{eq8}
\ee
a mere $1\sigma$ discrepancy.

By the end of the Moriond workshop, it was being reported \cite{r51} that $\sigma [e^+ e^- \ra$ Hadrons] is somewhat larger than previously thought, with the consequences that $a_\mu^{SM}(e^+ e^-)$ is somewhat larger as well. This reduces the discrepancy between experiment and the SM in \Eq{7}. At Lepton Photon 2003, in August 2003, P. Gambino \cite{r52} quoted a new value of the hadronic vacuum polarization contribution to $a_\mu$, based on a reanalysis by the Cryogenic Magnetic Detector experiment of its data, \cite{r53} and a new analysis by Hagiwara {\it et al.} The new value quoted by Gambino reduces the central value of the discrepancy $a_\mu^{\exp} - a_\mu^{SM}(e^+ e^-)$ in \Eq{7} from $35.6 \times 10^{-10}$ to $27.7 \times 10^{-10}$. If we may assume that the uncertainty in the discrepancy is still approximately as quoted in \Eq{7}, then the discrepancy has decreased to a 2 to 2.5$\,\sigma$ effect. In addition, the new $a_\mu^{SM}(e^+ e^-)$ is somewhat more consistent than the old one with $a_\mu^{SM}(\tau)$. \cite{r54}

Needless to say, it will be very interesting to see whether the discrepancy between the measured $a_\mu$ and the SM prediction becomes definitive or insignificant.

\section{Quark Flavor Physics and CP Violation}

Two of the most important purposes of quark flavor physics are to study the nature of CP violation, and to look for non-SM New Physics. New Physics can appear as ---
\begin{itemize}
\item A failure of the parameters in the CKM quark mixing matrix to describe all quark-sector weak decays, mixing, and CP violation,
\end{itemize} \vspace{-0.15in} or \vspace{-0.15in} \begin{itemize}
\item Inconsistent values when a SM quantity is measured in different ways,
\end{itemize} \vspace{-0.15in} or \vspace{-0.15in} \begin{itemize}
\item A non-SM rate for a rare decay,
\end{itemize} \vspace{-0.15in} or \vspace{-0.15in} \begin{itemize}
\item A non-SM CP-violating asymmetry in some decay.
\end{itemize}

\begin{figure}[hbt]
\begin{center}
\psfig{figure=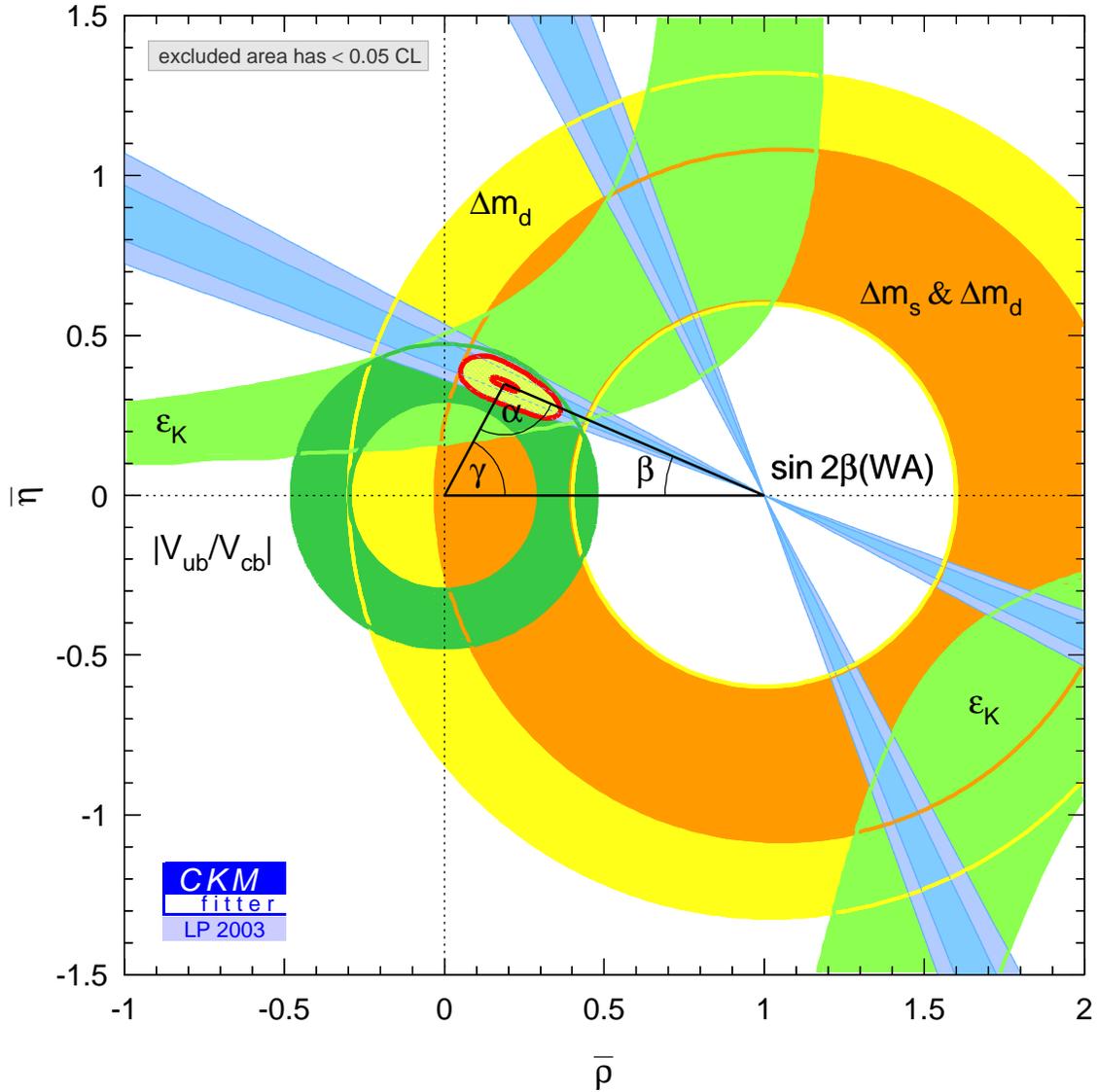,width=15cm}
\caption{Non-CP-violating and CP-violating experimental constraints on the parameters $\rho$ and $\eta$ in the CKM matrix $V$.
\label{f4}}
\end{center}
\end{figure}

In Wolfenstein's parametrization, the CKM matrix $V$ is described by the four parameters $\lambda, \,A, \,\rho$, and $\eta$. The parameters $\lambda$ and $A$ are fixed by $|V_{us}|$ and $|V_{cb}|$. Then, assuming that there is no New Physics in any decay, mixing, or CP violation, one obtains the constraints shown in Fig.~\ref{f4} \cite{r55} on the remaining parameters, $\rho$ and $\eta$. The interesting question is whether these constraints, and others yet to come, will prove to be consistent with each other as experimental and theoretical uncertainties shrink. (The axes of Fig.~\ref{f4} are $\bar{\rho} \equiv \rho (1 - \lambda^2 / 2)$ and $\bar{\eta} \equiv \eta (1 - \lambda^2 / 2)$. The factor $(1 - \lambda^2 / 2) \simeq 0.976$ is but a slight correction.)

\subsection{Non-CP-Violating Constraints}

Since $|V_{td}|^2 \propto (1 - \rho)^2 + \eta^2$, an accurate knowledge of $|V_{td}|$ would be a very significant constraint on $\rho$ and $\eta$. If there is no New Physics in $B_d-\overline{B_d}$ mixing, then $\Delta m_d$, the splitting between the masses of the mass eigenstates of the $B_d-\overline{B_d}$ system, is proportional to $|V_{td}|^2$. However, there is a large QCD uncertainty in the constant of proportionality. Thus, while we have learned that~\cite{r56}
\be
\Delta m_d = (0.500 \pm 0.008 \pm 0.006)\,\mathrm{ps}^{-1}~~,
\label{eq9}
\ee
this accurate value yields only a rather uncertain value of $|V_{td}|^2$, and consequently only a rather loose constraint on $\rho$ and $\eta$. This is the constraint indicated by the thick yellow doughnut-shaped region labelled $\Delta m_d$ in Fig.~\ref{f4}. Note that the inner boundary of this region coincides with that of the region labelled $\Delta m_s \:\& \:\Delta m_d$.

If neither $B_d-\overline{B_d}$ nor $B_s-\overline{B_s}$ mixing involves New Physics, then $|V_{td}|$ can be determined with reduced QCD uncertainty by using the relation
\be
\frac{\Delta m_d}{\Delta m_s} = \frac{|V_{td}|^2}{|V_{ts}|^2} \, \frac{1}{\xi^2_{\mathrm{QCD}}}\, \frac{m_d}{m_s}~~.
\label{eq10}
\ee
Here, $\Delta m_s$ is the splitting between the masses of the mass eigenstates of the $B_s-\overline{B_s}$ system, and $m_d$ and $m_s$ are, respectively, the average masses of the $B_d$ and $B_s$ mass eigenstates. The quantity $\xi_{\mathrm{QCD}}$ is a QCD correction.

Among the quantities in \Eq{10}, $\Delta m_d,\; m_d$, and $m_s$ are already well measured. By the unitarity of the CKM matrix, $|V_{ts}|^2 \cong |V_{cb}|^2$, and it is reported that \cite{r57}
\be
|V_{cb}| = (4.04 \pm 0.09 \pm 0.05 \pm 0.08) \times 10^{-2}~~.
\label{eq11}
\ee
The QCD correction $\xi_{\mathrm{QCD}}$ is calculated on the lattice, and at this workshop it was reported that~\cite{r58}
\be
\xi_{\mathrm{QCD}} = 1.21 (4)\,(5)~~.
\label{eq12}
\ee
Thus, aside from $|V_{td}|$, the only unknown in the relation of \Eq{10} is $\Delta m_s$. This quantity will be measured at the Tevatron, \cite{r59,r60} and then we will have a reasonably accurate determination of $|V_{td}|$.

(At present we have only a lower bound on $\Delta m_s$, leading via \Eq{10} to an upper bound on $|V_{td}|^2$. Combined with the loose information on $|V_{td}|$ coming from $\Delta m_d$ alone, this constrains $(\rho, \eta)$ to the orange doughnut-shaped region labelled $\Delta m_s$ and $\Delta m_d$ in Fig.~\ref{f4}.)

Since $|V_{ub}|^2 \propto \rho^2 + \eta^2$, measurements of $|V_{ub}|$ also serve to constrain $\rho$ and $\eta$. Several $|V_{ub}|$ measurements were reported at the workshop. \cite{r57,r56,r61} With the information available at Lepton Photon 2003, the $|V_{ub}|$ measurements constrain $(\rho, \eta)$ to the small green doughnut-shaped region centered about $(\rho, \eta) = (0,0)$ and labelled $|V_{ub}/V_{cb}|$ in Fig.~\ref{f4}.

\subsection{CP-Violating Constraints}

The SM description of CP violation leads us to expect {\em direct} CP violation (CP violation in decay amplitudes, rather than just in mixing amplitudes) in both $K$ and $B$ decays. This direct CP violation is difficult to predict quantitatively, especially in $K$ decays. However, it is gratifying that experiment has confirmed the qualitative expectation that $K$ decay should exhibit some nonvanishing degree of direct CP violation. This confirmation is the observation that $(\epp/\epsilon)_K$, a measure of direct CP violation, is nonzero. After the carrying out of very challenging measurements, we have
\be
\left( \frac{\epp}{\epsilon} \right)_K = \left\{
	\begin{array}{ll}
	(14.7 \pm 2.2) \times 10^{-4} &; \mbox{  NA48 \cite{r62}}  \nonumber \\
	(20.7 \pm 2.8) \times 10^{-4} &; \mbox{  KTeV \cite{r63}}
	\end{array} \right.
\label{eq13}
\ee

\begin{figure}[htb]
\begin{center}
\psfig{figure=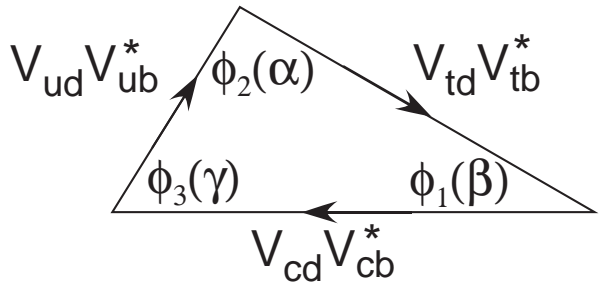,height=1.5in}
\caption{The CKM unitarity triangle.
\label{f5}}
\end{center}
\end{figure}

The SM description of CP violation leads us to expect {\em large} CP-violating effects in $B$ decays. The violation of CP in the $B$ system was discussed extensively at the workshop. \cite{r64}$^-$\cite{r70} Some of the tests of the SM that can be performed in this system may be nicely pictured in terms of the triangle with interior angles $\alpha,\; \beta$, and $\gamma$ in Fig.~\ref{f4}. \cite{r68} This triangle, known as the ``unitarity triangle'', is reproduced (with a change in scale) in Fig.~\ref{f5}. It is a pictorial representation in the complex 
plane of the CKM unitarity constraint $V_{ud}V_{ub}^* + V_{cd}V_{cb}^* + V_{td}V_{tb}^* = 0$. Its interior angles, being the phases of various products of CKM elements, are CP-violating quantities. As shown in Fig.~\ref{f5}, the angles $\alpha,\; \beta$, and $\gamma$ are also known, respectively, as $\phi_2,\; \phi_1$, and $\phi_3$. When the elements of $V$ are expressed in terms of $\lambda,\; A,\; \rho$, and $\eta$, and the triangle of Fig.~\ref{f5} is rescaled by dividing all its sides by $|V_{cd}V_{cb}^*|$, it becomes the triangle in Fig.~\ref{f4}.

To test --- and hopefully find a failing in --- the SM, one would like to use the CP-violating asymmetries in different $B$ decay modes to determine each angle in the unitarity triangle in several ways. Do all ways of measuring a given angle give the same answer? One would also like to measure the three different angles $\phi_2(\beta),\; \phi_1(\alpha)$, and $\phi_3(\gamma)$ independently. Do the measured values satisfy $\phi_1 + \phi_2 + \phi_3 = \pi$? And do the values obtained for these angles from CP-violating asymmetries agree with those inferred from the measured lengths of the sides of the unitarity triangle?

A very interesting start has been made on this program. From the CP-violating asymmetry in $\optbar{B}_d \ra \Psi K_S$ and very closely-related decay modes, it has been found that \cite{r71,r66}
\be
\sin 2\beta = 0.736 \pm 0.049~~.
\label{eq14}\
\ee
Thus, $\beta$ is in one of the four thin blue wedges emanating from the point $(\bar{\rho},\bar{\eta}) = (1,0)$ in Fig.~\ref{f4}. Now, the length of the side of the unitarity triangle running between this point and $(\bar{\rho}, \bar{\eta}) = (0,0)$ is unity by definition.  The length of the side opposite to $\beta,\; |V_{ud}V_{ub}| \cong |V_{ub}|$, is constrained by the ``$|V_{ub} / V_{cb}|$'' doughnut-shaped region in Fig.~\ref{f4}. The length of the remaining side, $|V_{td}V_{tb}| \cong |V_{td}|$, is constrained by the ``$\Delta m_s\: \& \:\Delta m_d$'' doughnut-shaped region. As we see from Fig.~\ref{f4}, one of the possible values of $\beta$ corresponding to \Eq{14}, $\sim 24^\circ$, is beautifully consistent with the range required by the intersection of the $|V_{ub} / V_{cd}|$ and $\Delta m_s \:\& \:\Delta m_d$ doughnut-shaped regions. As we also see, the constraint from the CP-violating parameter $\epsilon_K$ in the kaon system is consistent with everything else. Thus, as far as the measurements reflected in Fig.~\ref{f4} are concerned, the SM remains triumphantly successful.

\begin{figure}[htb]
\begin{center}
\psfig{figure=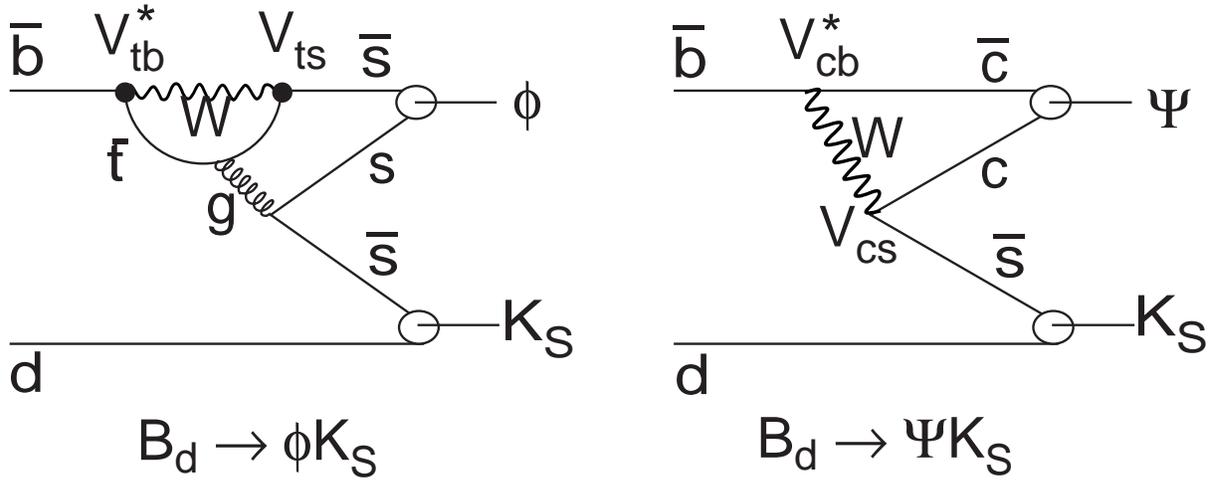,width=16cm}
\caption{The diagrams for the decays $B_d \ra \phi K_S$ and $B_d \ra \Psi K_S$.
\label{f6}}
\end{center}
\end{figure}

What happens if we compare the CP-violating asymmetry in $\optbar{B}_d \ra \phi K_S$ with that in $\optbar{B}_d \ra \Psi K_S$? Both reactions involve the same $B_d \leftrightarrow \overline{B_d}$ mixing, whose amplitude in the SM has phase $2\beta$. In the SM, the Feynman diagrams driving the two decays are the ones shown in Fig.~\ref{f6}.
As we see, the decay diagram for $B_d \ra \phi K_S$ is proportional to $V_{tb}^* V_{ts}$, while that for $B_d \ra \Psi K_S$ is proportional to $V_{cb}^* V_{cs}$. These two CKM  products are both real to a very good approximation. Thus, in the SM, the CP-violating asymmetries in $\optbar{B}_d \ra \phi K_S$ and $\optbar{B}_d \ra \Psi K_S$ should both yield the same CP-violating parameter, $\sin 2\beta$. Do they give the same value for this parameter?

At the time of the Moriond workshop, there was a hint that perhaps the values of ``$\sin 2\beta$'' extracted from $\optbar{B}_d \ra \phi K_S$ and $\optbar{B}_d \ra \Psi K_S$ do {\em not} agree, with $\optbar{B}_d \ra \phi K_S$ yielding a ``$\sin 2\beta$'' whose central value is actually negative, in disagreement with \Eq{14}, but with a large uncertainty. \cite {r66,r64} There was discussion of the possibility that supersymmetry could modify $\optbar{B}_d \ra \phi K_S$ so as to make the ``$\sin 2\beta$'' inferred from this decay neglecting supersymmetry negative. \cite{r72} By the time of Lepton Photon 2003, the value of ``$\sin 2\beta$'' extracted from $\optbar{B}_d \ra \phi K_S$ assuming the SM describes this decay was being reported as \cite{r73}
\be
\qsin2b{\phi} = \left\{
	\begin{array}{ll}
	+0.45 \pm 0.43 \pm 0.07 &; \;\; \mathrm{BaBar}  \nonumber \\
	-0.96 \pm 0.50 \phantom{00}_{-0.11}^{+0.09} & ;\;\; \mathrm{Belle}
	\end{array} \right.
\label{eq15}
\ee
Obviously, the BaBar and Belle measurements are not terribly consistent with each other. The Belle value of $\qsin2b{\phi}$ is more than $3\,\sigma$ from the BaBar--Belle average value of $\left. \sin 2\beta \right| _{\mathrm{From}\:\Psi K_S}$, \Eq{14}. But the BaBar value of $\qsin2b{\phi}$ is perfectly consistent with $\left. \sin 2\beta \right| _{\mathrm{From}\:\Psi K_S}$. It will be very interesting to see how this situation evolves.

We do not yet have very definitive information on the angles $\alpha$ and $\gamma$. CP-violating asymmetries in $B \ra \pi^+\pi^-$ bear on $\alpha$. Two asymmetries, $S_{\pi\pi}$ and $A_{\pi\pi}$, which must satisfy $S^2_{\pi\pi} + A^2_{\pi\pi} \leq 1$, \cite{r64} are being probed in this decay mode. At Moriond, it was reported that
\begin{center} \begin{tabular}{ccc} 
\underline{$S_{\pi\pi}$}       &  \underline{$A_{\pi\pi}$} 
	& \underline{Experiment} \\
\vspace{-0.12in} \\ 
$-1.23\pm 0.41\phantom{00}^{+0.08}_{-0.07}$ & $+0.77\pm 0.27\pm 0.08$ 
	& Belle \cite{r65} \\
$+0.02 \pm 0.34 \pm 0.05$      & $+0.30\pm 0.25\pm 0.04$ & BaBar \cite{r66} 
\end{tabular} \end{center} 
\vspace{0.2cm} 
At Lepton Photon 2003, new results were reported for BaBar:
\begin{center} \begin{tabular}{ccc} 
\underline{$S_{\pi\pi}$}       &  \underline{$A_{\pi\pi}$} 
	& \underline{Experiment} \\
\vspace{-0.12in} \\
$-0.40 \pm 0.22 \pm 0.03$      & $+0.19\pm 0.19\pm 0.05$ & BaBar \cite{r74} 
\end{tabular} \end{center}
\vspace{0.2cm} 
Clearly, the uncertainties in $S_{\pi\pi}$ amd $A_{\pi\pi}$ are still large. Once these quantities are pinned down, the determination of $\alpha$ will require study of the additional difficult decay mode $B \ra \pi^0 \pi^0$. Alternatively, one can try to determine $\alpha$ from the $B \ra \pi^+ \pi^-$ observables, without input from $B \ra \pi^0 \pi^0$, by using the pure-penguin process $B^+ \ra K^0 \pi^+$ plus SU(3) symmetry to obtain the magnitude of the penguin-diagram contribution to $B \ra \pi^+ \pi^-$ and $B \ra \pi \ell \nu$ plus factorization to obtain the magnitude of the tree-diagram contribution. \cite{r64}

A variety of approaches to measuring $\gamma$ were discussed at the workshop. These included trying to extract $\sin (2\beta + \gamma)$ from $B \ra D^{*^-} \pi^+$, \cite{r68} using ratios of rates for B decay into various $K\pi$ final states, \cite{r64} and other approaches. \cite{r69}

\subsection{Rare Decays}

\begin{figure}[htb]
\begin{center}
\psfig{figure=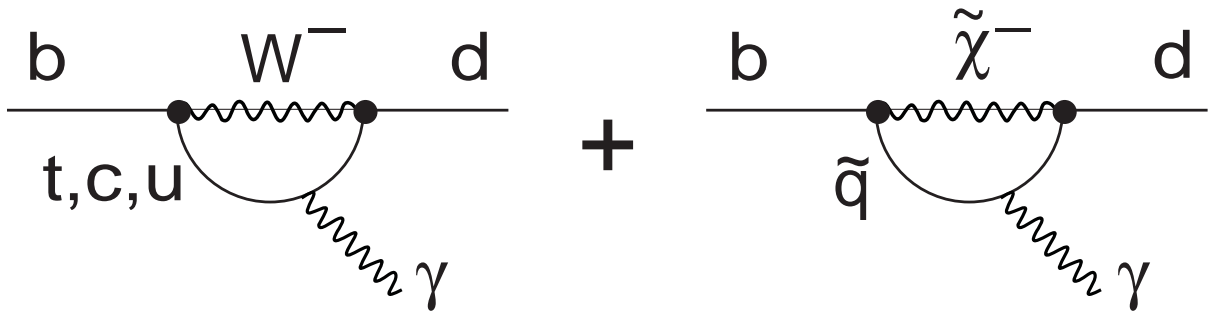,width=16cm}
\caption{A SM diagram, and one from supersymmetry, for $b \ra d\gamma$. In the latter diagram, $\tilde{\chi}^-$ is a chargino and $\tilde{q}$ a squark.
\label{f7}}
\end{center}
\end{figure}

Another way to seek evidence of New Physics is to measure the rates for the decays $b \ra s\gamma,\;b \ra d\gamma,\; b \ra s\ell^+\ell^-$, and $b \ra d\ell^+\ell^-$. \cite{r75,r72,r76} These are all rare decays that occur at the loop level, rather than the tree level, in the SM. As a result of this SM suppression, these are decays in which small effects from New Physics are potentially visible. An example is the decay $b \ra d\gamma$, which can receive contributions from both the SM diagram on the left of Fig.~\ref{f7}, and the supersymmetric one on the right.

Experimentally, both exclusive and inclusive decay modes are studied.There are new, tighter limits on several so-far unobserved modes, and we may be close to observing a first $b \ra d\gamma$ signal. \cite{r75} It is reported that BR[$B \ra K\ell^+\ell^-$] is [$4.8\,^{+1.0}_{-0.9} \pm 0.3 \pm 0.1$ (model uncertainty)] $\times 10^{-7}$ according to Belle, and $[6.9 \,^{+1.5}_{-1.3} \pm 0.6] \times 10^{-7}$ according to BaBar. \cite{r77} By comparison, in the SM, BR[$B \ra K\ell^+\ell^-] = (3.5 \pm 1.3) \times 10^{-7}$. \cite{r78} While this prediction is somewhat below the measured values, the discrepancy is not conclusive.

\subsection{Anomalous CP-Violating Asymmetries}

When the predicted SM CP-violating asymmetry in some decay mode is very small, the observation of a much larger asymmetry in that mode would signal the presence of New Physics. An interesting accessible mode where the predicted SM asymmetry is indeed small --- of order 2\%~--- is $B_s \ra \Psi\phi$. This decay can be studied at hadron colliders, and has already been observed by both the CDF \cite{r59} and D0 \cite{r79} detectors at the Tevatron. We look forward to the eventual study of the CP asymmetry.

\section{Collider Physics}

We regret that, because at Moriond the collider electroweak results were presented not long before the summary session, we were not able to include them in any detail. We can make only a few comments:

Early results from Run II of the Tevatron, obtained with an upgraded collider complex and upgraded detectors, were reported at the workshop. \cite{r80,r81} Further results were reported at Lepton Photon 2003, in August 2003. We give just a very few examples. Various measurements of the top quark production cross section, \cite{r82} based on successful observation of a variety of top decay channels, show that the many components of the upgraded CDF and D0 detectors are working well. The D0 detector has produced the tightest existing bound, $M_s > 1.38$ TeV, on the scale $M_s$ of Kaluza-Klein gravitons in large extra dimensions. \cite{r45} The CDF detector has produced the world's best measurements of the $B_s$ and $\Lambda_b$ masses: $m(B_s) = (5365.50 \pm 1.60)$ MeV and $m(\Lambda_b) = (5620.4 \pm 2.0)$ MeV. \cite{r79} We look forward to the many new results that the Tevatron will yield. \cite{r83}

\section{Conclusion}

Exciting, interesting, and intriguing results were presented at the 2003 Moriond Workshop on Electroweak Interactions and Unified Theories. In addition, there was a particularly illuminating and stimulating exchange of ideas.

\section*{Acknowledgments}
It is a pleasure to thank J. Tran Thanh Van, the creator of the Moriond workshops, the Program Committee, the speakers, and the participants for a very fruitful meeting. The author is grateful to F. Montanet for his crucial collaboration in the preparation of the talk on which this manuscript is based, and to Susan Kayser for her essential role in the production of the manuscript. This work was supported by Fermilab, which is operated by URA under DOE contract No. DE-AC02-76CH03000.

\end{document}